\documentclass[aps,twocolumn,showpacs]{revtex4}
\usepackage{epsfig,amsmath,psfrag,graphicx,dsfont,mathrsfs,subfigure}

\newcommand\ket[1]{\left|#1\right\rangle}
\newcommand\bra[1]{\left\langle#1\right|} 
\newcommand\braket[2]{ \langle #1 | #2 \rangle }
\newcommand\ketbra[2]{ | #1 \rangle\!\langle #2 | }

\newcommand{\eps}{\mathcal E} \newcommand{\id}{\mathds 1}
\newcommand{\ten}{\otimes} 
\newcommand{\tr}{\mbox{tr}} \renewcommand{\rho}{\varrho}
\newcommand{\sys}{\mathcal S} \newcommand{\res}{\mathcal R}

\begin{document}

\title{Characterization of Decoherence from an Environmental Perspective}
\author{Julius Helm}
\author{Walter T. Strunz}
\affiliation{Institut f\"{u}r Theoretische Physik, Technische Universit\"{a}t Dresden, 
01062 Dresden, Germany}
\author{Stephan Rietzler}
\affiliation{B\"{o}hmeweg 55, 89075 Ulm, Germany}
\author{Lars Erik W\"{u}rflinger}
\affiliation{Institut de Ci\`{e}nces Fot\`{o}niques, 08860 Castelldefels (Barcelona), Spain}

\date{\today}
\begin{abstract}
  For the case of phase damping (pure decoherence) we investigate the
  extent to which environmental traits are imprinted on an open
  quantum system. The dynamics is described using the quantum channel
  approach. We study what the knowledge of the channel may reveal
  about the nature of its underlying dynamics and, conversely, what
  the dynamics tells us about how to consistently model the
  environment. We find that for a Markov phase-damping channel, that
  is, a channel compatible with a time-continuous Markovian evolution,
  the environment may adequately be represented by a mixture of only a
  few coherent states. For arbitrary Hilbert space dimension $N\geq 4$
  we refine the idea of {\it quantum phase damping}, of which we show
  a means of identification. Symmetry considerations are used to
  identify decoherence-free subspaces of the system.
\end{abstract}
\maketitle

\section{Introduction}

Decoherence describes the loss of characteristic traits of quantum
theory. For the success of emerging quantum technologies a detailed
understanding of decoherence is of great relevance. Schemes to avoid
and counter its effects need to be developed. Besides, decoherence
offers insight into the much-debated quantum-to-classical transition
\cite{Giulini, ZurekReview, StrunzDecoherence}.  The microscopic
dynamics leading to decoherence might be based on very diverse
grounds, reaching from purely classical phase kicks to a quantum
mechanical formulation based on coupling the system of interest to
some quantum environment.  Hence, a further characterization of
different microscopic mechanisms leading to decoherence is desirable.

Phase damping (or dephasing) denotes the case of pure decoherence,
corroding the coherences of a quantum state while leaving the
probabilities, that is, the diagonal elements of the density matrix,
intact. The dissipation-less transition of a pure state into a
classical mixture when described in the basis of energy eigenstates
may serve as an example. Despite its simple nature, phase damping is
enough to completely disentangle quantum states \cite{Yu2003}.

For weak system-environment coupling and short environmental
correlation times decoherence may be modelled in terms of Markovian
dynamics \cite{BreuerBook}.  Here, the future evolution depends solely
on the system's present state, rather than on anterior times.  Yet,
there are of course instances where this approximation is not valid.
Given the dynamics of a quantum system it would be valuable to have a
means of deciding whether the dynamics is Markovian or not.  This
point has been studied lately both in a continuous approach based on
the information flow between system and environment \cite{Breuer2009},
as well as from a snapshot point of view \cite{Wolf2008a,Wolf2008b},
where the continuous dynamics is by construction unavailable. Rather,
the state of the quantum system is known at separate times, only.

Another interesting question is whether the phase damping is due to
coupling to a ``real'' quantum-mechanical environment, or wether it
can equally be explained in terms of stochastically fluctuating,
classical fields \cite{AlickiLendi,NielsenChuang}. The latter is a
convex combination of unitary transformations, that is, random unitary
(RU) dynamics.  While phase damping of a single qubit or qutrit may
always be described as RU dynamics, in Hilbert spaces of dimension $N
\geq 4$ one cannot always find such a representation
\cite{LandauStreater, DAriano, HelmStrunz}.

In the article at hand we study the characteristics of phase damping
from an environmental point of view. Phase damping is described
utilizing the overlap of {\it dynamical vectors} relative to the {\it
  phase damping basis}. The nature of the dynamics is reflected by the
set of dynamical vectors, or, conversely, the properties of the
dynamical vectors determine the dynamics to a certain extent. In this
context, we show that in case of Markovian phase damping the dynamical
vectors can be identified with coherent states. Likewise, we give
instructions for a physical model of ``quantum phase damping'' for
arbitrary Hilbert space dimension $N \geq 4$, that is, phase damping
which does not allow for a RU representation.

The article is structured as follows. Section \ref{sec:channels}
overviews the theoretical background and serves as an introduction to
the formal notation. In Sec.~\ref{sec:singlequbit} we exemplary study
phase damping on a single qubit, where all characteristics introduced
so far actually coincide. Sections \ref{sec:generalmarkovian} and
\ref{sec:randomunitary} address the Markovianity and the possibility
of finding a RU representation, respectively. In Sec.~\ref{sec:DFS} we
discuss the appearence of decoherence-free subspaces due to symmetries
in our formalism.

\section{Quantum Channels}
\label{sec:channels}
Based on the fundamental assumption of no initial correlations between
the system $\rho$ and its environment, the most general quantum
evolution is given by a completely positive map $\eps: \rho \mapsto
\eps(\rho)$. In a Hilbert space of dimension $N$, these maps (or
``quantum channels'') can always be written in terms of at most $N^2$
{\it Kraus operators} $K_i$ such that
\begin{equation}
  \label{eq:kraus}
  \rho \mapsto \rho'= \eps(\rho)=\sum_i K_i \rho K_i^\dagger
\end{equation}
(here and in the following we denote the initial state by $\rho$ and
its map by $\rho'$).  It is usually assumed that the map is
trace-preserving, $\sum_i K_i^\dagger K_i = \mathds 1$, so as to
preserve probability. If, in addition, the completely mixed state is
mapped onto itself: $\sum_i K_i K_i^\dagger = \mathds 1$, the channel
is said to be unital or doubly stochastic
\cite{ZyczkowskiBook}. Throughout the article we will assume that
$\rho$ and $\rho'$ live in the same Hilbert space, that is, the
channel $\eps$ maps the set of states on a Hilbert space of dimension
$N$ onto itself.

When considering a quantum channel of form (\ref{eq:kraus}), no
particular assumptions are made about the nature of the underlying
continuous dynamics.  Rather, only a snapshot of the quantum system at
a given time is revealed.  Nevertheless, in some cases it is possible
to gather information about the nature of the physical processes
involved. In the remainder of this section, we want to discuss how
certain additional assumptions about the structure of the channel may
set restrictions to the underlying dynamics or vice versa.

\subsection*{Markovian Channels}
A quantum channel $\eps$ is said to be Markovian, if there exists a
generator $\mathcal L$ of a quantum dynamical semigroup and a time
$t>0$ such that
\begin{eqnarray}
  \label{eq:3}
  \eps(\rho) = e^{\mathcal L t} \rho 
\end{eqnarray}
\cite{Wolf2008a,Wolf2008b}. That is, the channel may be understood as
a snapshot of a time-continuous Markovian evolution. The generator
$\mathcal L$ may be written in {\it Lindblad} form \cite{BreuerBook}
\begin{eqnarray}
  \label{eq:lindblad}
  \mathcal L(\rho) = -i [H,\rho]+ \frac{1}{2}\sum_{i = 1}^r 
  \left\{
    \left[ L_i^{} \rho,L_i^\dagger\right] +
    \left[ L_i^{}, \rho L_i^\dagger\right]
  \right\}. 
\end{eqnarray}

The Markov property of a channel is closely related to the notion of
{\it infinite divisibility} \cite{Denisov,Wolf2008a}. A channel $\eps$
is called infinitely divisible if, for all $\nu \in \mathds N$, there
exists a channel $\eps_\nu$ with $(\eps_\nu)^\nu = \eps$. Surely, a
Markov channel $\eps$ is infinitely divisible: for any given $\nu \in
\mathds N$ it can be written as $\nu$-fold concatenation of the
channels $\eps_{\nu} = e^{\mathcal L t/\nu}$. The converse statement,
however, is not true in general \cite{Denisov}.

\subsection*{RU Channels}
One of the standard approaches to the quantum channel formalism is
based on the reduced dynamics of a system interacting with its
environment \cite{NielsenChuang}. In this context, decoherence of an
open quantum system is inevitably linked to growing entanglement
between system and environment \cite{ZurekReview}. Yet, there are instances
of irreversible dynamics that may be modeled entirely without invoking
a quantum environment at all. An important example is given by RU
dynamics, where the quantum channel may be written as a convex
combination of unitary transformations
\begin{eqnarray*}
  \eps(\rho) = \sum_i p_i U_i \rho U_i^\dagger \quad\quad 
  \big(p_i > 0, \sum_i p_i =1\big). 
\end{eqnarray*}
The dynamics may thus be thought of as originating from classical
fluctuations, hence also termed ``random external fields''
\cite{AlickiLendi, NielsenChuang}. It is known, for example, that for
a single qubit all doubly stochastic channels are of RU type
\cite{LandauStreater}. These RU channels gain some significance in the
field of quantum error correction, where they stand out due to the
fact that they may be undone completely \cite{GregorattiWerner}.  More
recently they have also been applied to quantum networks \cite{Alber}.

\subsection*{Phase-Damping Channels}
Phase-damping channels are among the simplest conceivable quantum
channels. They are defined by the requirement that in a given basis
$\{|n\rangle\}$---the {\it phase-damping basis}---no population
transfer takes place. The only effect of the ``environment'' is thus
to change coherences $\langle n | \rho |m\rangle$ with $n\neq m$ and
to leave all $\langle n | \rho |n\rangle$ with $n=1,\ldots,N$
untouched. In other words, the projectors are constants of motion:
$\eps(\ketbra{n}{n}) = \ketbra{n}{n}$ for all $n$.

We conclude that the Kraus operators have to be diagonal in this
basis, $K_i=$diag$(a_{i1}, a_{i2}, \ldots , a_{iN})$ and,
correspondingly, the whole map $\eps$ is diagonal, too. We find
\begin{equation}
  \label{eq:dynamicalvectors}
  \rho'_{mn} = \langle a_n|a_m\rangle \rho_{mn}
\end{equation}
with $\{ a_n=(a_{1n},a_{2n},\ldots,a_{rn})\}$ any set of $N$
normalized complex vectors. It is then sometimes convenient to
introduce the matrix $D$ with $D_{mn} = \langle a_n|a_m\rangle$ to
write the phase-damping channel in the short form $\rho' = D \star
\rho$, where $\star$ is the {\it Hadamard product}, that is, the
entry-wise product of matrices of the same size: $\rho'_{mn} = D_{mn}
\rho_{mn}$ \cite{Havel}. From these considerations it is clear that
phase-damping channels are among the doubly stochastic channels.

If the quantum channel $\eps$ is defined via the system's coupling to
a quantum mechanical environment, the vectors $\ket{a_n}$ may be seen
as relative states of the environment, that is, relative to the states
of the distinguished basis (see also Sec.~\ref{sec:randomunitary}).
Then the overlap $\braket{a_n}{a_m}$, seen as a function of time, may
be related to studies of fidelity decay \cite{2004GorinStrunz}.  Yet,
this relative state picture need not hold in general: the case of RU
dynamics shows that in certain circumstances decoherence may be
attributed to stochastic, fluctuating ``classical'' fields.

In many situations the dynamical vectors $\ket{a_n}$ are of course
unknown a priori. In particular this holds true in an experimental
setup where the matrix $D$ is acquired via quantum process tomography
\cite{NielsenChuang}. One way of obtaining dynamical vectors
$\ket{a_n}$ from $D$ is by using the Cholesky factorization
\cite{Gentle}. Given the non-negative matrix $D$, the Cholesky
factorization gives $D = L L^\dagger$, with $L$ a lower triangular
matrix ($L$ is in general not unique). The $n$-th row of $L$ may then
be identified with a complex vector $\ket{a_n} \in \mathds C^d$ such
that $D_{mn} = \braket{a_m}{a_n}$. If $D$ is a positive semi-definite
matrix of rank $r<d$, there exists a unique $L$ with columns $r+1$
through $d$ identical to zero \cite{Gentle}. That is, the vectors
$\ket{a_n}$ may be chosen as elements of $\mathds C^r$. In the
following sections we will study what these dynamical vectors
$\ket{a_n}$ reveal about the nature of the underlying dynamics.

\section{The Single Qubit Case}
\label{sec:singlequbit}
Without revealing too much about the details we want to state some
results of the following sections. The case of a single qubit stands
out due to the fact that a phase-damping channel is always Markovian
(i.e., a snapshot of Markovian dynamics) and it is of RU type. These
findings of course do not allow for generalization to higher
dimensional systems, yet they have some potential for building
intuition.  For a more rigorous approach as well as some missing
definitions see Secs. \ref{sec:generalmarkovian} and
\ref{sec:randomunitary}.

For a single qubit the phase-damping map is defined by the matrix
\begin{equation}
  \label{eq:Dqubit}
  D = \left(\begin{array}{cc} 1 & \langle a_2|a_1\rangle \\
      \langle a_1|a_2\rangle & 1\end{array}\right).
\end{equation}
Thus, a single complex number $\langle a_2|a_1\rangle =: c$ with
modulus less than one determines the most general single-qubit phase-
damping channel.  Infinite divisibility of a phase-damping channel has
to be formulated in terms of the Hadamard product (see also
Sec.~\ref{sec:generalmarkovian}), that is, the matrix $D_\nu$ with
$(D_\nu)_{mn} = (D_{mn})^{1/\nu}$ has to be checked as to its
positivity.  It is quite straightforward to see that the matrix $D$ in
Eq.~(\ref{eq:Dqubit}) passes this test, which lets us conclude that a
single qubit phase-damping channel is always Markovian (see also
Sec.~\ref{sec:generalmarkovian}).

Another remarkable feature of single qubit phase damping---which we
will later show to be intimately connected to Markovianity---is that
the dynamical vectors in (\ref{eq:dynamicalvectors}) may be chosen
from the set of coherent states $\{ \ket{\alpha} | \alpha \in \mathds
C \}$ of a harmonic oscillator. These are eigenstates of the
annihilation operator, $a \ket \alpha = \alpha \ket \alpha$, and may
be seen as displaced vacuum states: $\ket \alpha = e^{\alpha
  a^\dagger- \alpha^* a} \ket 0$ \cite{ScullyBook}. In order to see
this, note that for $c \ne 0$ we may simply let
\begin{eqnarray}
  \label{eq:6}
  c =: e^{-2 \gamma - i \omega}
\end{eqnarray}
with $\gamma \in \mathds R_+$ and $\omega \in [0, 2 \pi)$.  We then
define the two-mode coherent states $\ket{\alpha_n} := e^{-i \omega_n}
\ket{\sqrt{\gamma} l_n}$, where $l_1=(1,1), l_2=(1,-1)$ and $\omega_1
= -\omega_2 = \omega/2$.  It is easy to see that these states give the
correct overlap, that is, $\braket{\alpha_2}{\alpha_1} = c$.  In this
vein we can thus always define the channel in terms of coherent states
$\ket{\alpha_n}, n=1,2$, leading to Markovian dynamcis. Written in
Lindblad form the master equation attains its well-known form
\begin{eqnarray}
  \label{eq:8}
  \mathcal L(\rho) = -i \frac{\omega}{2}[\sigma_z,\rho] - 
  \frac{\gamma}{2} [\sigma_z,[\sigma_z,\rho]], 
\end{eqnarray}
where channel (\ref{eq:Dqubit}) with $c$ from (\ref{eq:6}) is
obtained as a snapshot for $t=1$.

Alternatively, we may choose to write the overlap of states in the
form $\langle a_2|a_1\rangle=(2p-1)e^{-i\theta}$ with $0\le p\le 1$,
obtaining the common quantum channel representation
\cite{NielsenChuang}
\begin{equation}
  \label{singlequbit}
  \eps (\rho) = e^{-i\frac{\theta}{2}\sigma_z}
  \left(p\rho + (1-p)\sigma_z \rho \sigma_z\right) e^{i\frac{\theta}{2}\sigma_z}. 
\end{equation}
In this notation it is rather obvious, hence, that the channel is RU,
which is always true for a single qubit or qutrit
\cite{LandauStreater,DAriano}.

\section{Phase-damping Markov Processes and coherent states}
\label{sec:generalmarkovian}

As introduced in Sec.~\ref{sec:channels}, the mapping of a quantum
state subject to phase damping may be written using the Hadamard
product. Infinite divisibility is equivalent to positivity of the
matrices $D_\nu$, where $(D_\nu)^\nu = D_\nu \star \ldots \star D_\nu
= D$, $\nu \in \mathds N$, i.e., $(D_{\nu})_{mn} = (D_{mn})^{1/\nu}$.
While it is clear that every Markov channel is infinitely divisible,
note that the converse also holds in case of phase damping (when all
$D_{mn} \ne 0$).

The argument is based on a theorem by Denisov \cite{Denisov} which
states that an infinitely divisible channel $\eps$ is of the form
$\eps = e^{\mathcal L} E$, where $E$ is an idempotence with $E
\mathcal L E = \mathcal L E$ \cite{Denisov,Wolf2008a}.  In case of
phase damping, however, the diagonal character of the map together
with the relation $E\eps = E e^{\mathcal L} E = e^{\mathcal L} E =
\eps$ implies $E=\mathds 1$ already, whenever all $D_{mn} \ne
0$. Thus, Markovianity follows directly from infinite divisibility in
this case. Recall that this is certainly not true for channels in
general.

From Sec.~\ref{sec:singlequbit} we already know that any single-qubit
phase-damping channel (with $c\ne 0$) is infinitely divisible and
hence Markovian, but what about higher dimensions?  A simple example
shows that already for a $3$-dimensional quantum system positivity may
be violated: Let a $3$-state phase-damping channel be given by
\begin{eqnarray*}
  D = \left(
    \begin{array}{ccc} 
      1 & i \alpha & -i \alpha \\
      -i \alpha & 1 & \alpha \\
      i \alpha & \alpha & 1
    \end{array}
  \right), 
\end{eqnarray*}
with real $\alpha$.  Then for $\frac{1}{3} < \alpha \leq \frac{1}{2}$
the matrix $D$ is positive, but all (Hadamard) square roots $D^{1/2}$
have one negative eigenvalue equal to $1-\sqrt{3 \alpha}$. We have
thus found a single-qutrit phase-damping channel which may not be
identified as a snapshot of Markovian evolution.

The notion of infinite divisibility in this section implicitly assumes
all fractional powers of the initial channel to be phase damping and
therefore diagonal. This excludes the rather peculiar case where some
dynamics is phase damping for a particular time $t$ only, but may well
change populations at other times. Consider, for example, the unitary
one-qubit map $\eps(\rho) = U \rho U^\dagger$ with $U = \mbox{exp}(-i
\pi \sigma_x t)$, which is trivially phase damping for $t =
1,2,3,\ldots$.

From these considerations it is clear that the underlying generator
$\mathcal L$ of the Markov dynamics is diagonal: $\mathcal
L(\ketbra{m}{n}) = z_{mn} \ketbra{m}{n}$. This, in turn, assures the
diagonal character of both the Hamiltonian $H=:\sum_n w_n
\ketbra{n}{n}$ and the Lindblad operators $L_i=:\sum_n l_n^{(i)}
\ketbra{n}{n}$ (see App.~\ref{app:diagonality}), thereby leading to
the relation
\begin{eqnarray}
  \label{eq:generator}
  z_{mn} = -i(\omega_m - \omega_n) + 
  \braket{l_n}{l_m} - \frac{1}{2}(\|l_m\|^2 + \|l_n\|^2), 
\end{eqnarray}
where $\l_n := (l_n^{(1)},\ldots, l_n^{(r)})$, $r$ is the number of
Lindblad operators in (\ref{eq:lindblad}).

An M-mode coherent state may be written as a displacement of the
vacuum
\begin{eqnarray*}
  \ket{\alpha} = e^{-\|\alpha\|^2/2}\,e^{\alpha_1 a_1^\dagger}\ten
  \ldots \ten e^{\alpha_M a_M^\dagger} \;\ket{0}_1\ten\ldots\ten\ket{0}_M
\end{eqnarray*}
with bosonic creation operators $a_i^\dagger$ \cite{ScullyBook}. For
two coherent states $\ket{\alpha}, \ket{\beta}$ this leads to an
overlap
\begin{eqnarray}
  \label{eq:overlap}
  \braket{\beta}{\alpha} = 
  e^{\langle \beta , \alpha \rangle -\frac{1}{2}(\|\alpha\|^2+\|\beta\|^2)},
\end{eqnarray}
where $\langle \cdot , \cdot \rangle$ denotes the standard scalar
product in $\mathds C^M$ and $\| \alpha \|^2 = \langle \alpha,\alpha
\rangle$.

Based on comparison of Eqs. (\ref{eq:generator}) and
(\ref{eq:overlap}) we define $r$-mode coherent states of the following
form
\begin{eqnarray*}
  \ket{\alpha_n(t)} =
  e^{-i\omega_n t}\ket{l_n \, \sqrt t}, \quad n = 1, \ldots, N,
\end{eqnarray*}
and find that for Markovian phase damping we may define the channel in
terms of {\it coherent states} such that
\begin{eqnarray*}
  \label{eq:5}
  D_{mn} = e^{z_{mn} t}\big |_{t=1} 
  =\braket{\alpha_n (t)}{\alpha_m (t)} \, \big|_{t=1}.
\end{eqnarray*}
Any Markovian phase-damping channel $\eps$ may therefore be obtained
as the reduced dynamics of the system interacting with an environment
of harmonic oscillators, all in coherent states. At first sight, the
time dependence of the coherent states $\ket{\alpha_n(t)} = e^{-i
  \omega_n t} \ket{\sqrt{t} l_n}$ may seem quite queer. Yet, this
should not be too surprising given that a finite reservoir would
normally lead to memory effects. In order to preserve Markovianity the
dynamics thus has to be strongly driven, so as to prevent the
back-flow of information from the environment to the system
\cite{Breuer2009}. 

As a final remark we add that the $\sqrt{t}$-dependence of the
centroid of the environmental coherent states reflects the fundamental
relevance of standard Brownian motion for all (continuous) Markov
processes.

\section{RU vs. Quantum phase damping}
\label{sec:randomunitary}
We have seen in the previous section that for qutrits---or larger
systems---phase-damping channels need not be Markovian. In a similar
spirit, one may ask whether RU representations exists for any
dimension: can all phase-damping processes be written as a convex sum
of unitary maps? For a qubit, as for the question of Markovianity, the
answer is positive.  In general, however, the answer is no, as can be
found in \cite{LandauStreater}. For a two-qubit system, that is $N=4$,
a physical model for such a non-RU (or {\it quantum}) phase-damping
channel is described in Ref. \cite{HelmStrunz}.

Our aim here is to offer a method how to construct non-RU
phase-damping channels in arbitrary dimension, extending earlier
work. These serve as specific examples; it is an entirely different
and challenging matter to test a given phase-damping channel for the
RU property.

Our method of identification of such a quantum phase-damping channel
rests on extremality with respect to the convex set of doubly
stochastic channels. Due to a result by Landau and Streater it is
known that there exist non-unitary, extremal maps in the convex set of
diagonal doubly stochastic maps \cite{LandauStreater}.  Extremality is
guaranteed for channels where the projectors $\ketbra{a_n}{a_n}$
obtained from the dynamical vectors $\{\ket{a_1},\ldots,\ket{a_N}\}
\subset \mathds C^r$ in Eq. (\ref{eq:dynamicalvectors}) form a
(possibly overcomplete) operator basis on $\mathds C^r$. Note that
extremality requires $r^2 \leq N$ [remember that $r$ denotes the
number of operators used in Eq.~(\ref{eq:kraus}) or, likewise, the
dimensionality of the vectors $\ket{a_n}$, $N$ is the dimension of the
quantum system].

\begin{figure}[t]
  \includegraphics[width=.6\linewidth]{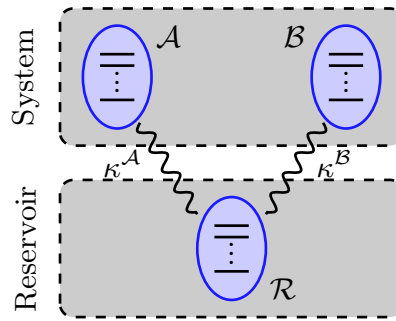}
  \caption{\small (Color online) The construction of an extremal
    phase-damping channel is based on a bipartite system of qudits
    $\mathcal A$ and $\mathcal B$, locally coupling (via
    $\kappa^{\mathcal A}$ and $\kappa^{\mathcal B}$) to a qudit
    reservoir $\mathcal R$.}
  \label{fig:figure1}
\end{figure}

The construction of the channel rests on a Hamiltonian $H$ locally
coupling two qudits (d-dimensional quantum systems) to a single qudit
environment (cf. Fig.~\ref{fig:figure1}). Then, by construction, $r^2
= d^2 = N$.  In the usual notation we set
\begin{eqnarray}
  \label{eq:hamiltonian}
  H = H_{\sys} + H_{\mathcal I} + H_{\res},  
\end{eqnarray}
where $H_{\sys}$ and $H_{\res}$ denote the Hamiltonian describing
system and reservoir, respectively. The local coupling of qudits
$\mathcal A$ and $\mathcal B$ to the reservoir $\mathcal R$ may be set
to
\begin{eqnarray*}
  H_{\mathcal I} = \sum_{i,j} 
  \left(
    \kappa^{\mathcal A}_{ij} \sigma^{\mathcal A}_i \ten \sigma^{\mathcal R}_j 
    + \kappa^{\mathcal B}_{ij} \sigma^{\mathcal B}_i \ten \sigma^{\mathcal R}_j 
  \right). 
\end{eqnarray*}
In order to invoke a phase-damping channel on the system we have to
require $H_{\sys}$ as well as all operators $\sigma^{\mathcal A}_i$,
$\sigma^{\mathcal B}_i$ to be diagonal (the $\sigma$-operators will be
specified below).

For any given time $t$ and assuming the usual product initial state,
$\rho \ten \sigma$, this dynamics leads to the phase-damping channel
\begin{eqnarray}
  \label{eq:tqd-channel}
  \eps_t (\rho) := 
  \rho' &=& \tr_\res \left( e^{-i H t}
    \left( \rho \ten \sigma \right) 
    e^{i H t} \right) \nonumber \\
  &=& \tr_\res \left(U (\rho \ten \sigma ) U^\dagger \right).
\end{eqnarray}
Due to the restriction to diagonal system Hamiltonian and diagonal
coupling, the unitary map $U$ allows for a diagonalization in the
phase-damping basis $\{\ket{n}\}$. The interaction may thus be
expressed in fashion of a controlled-unitary operation
\cite{ZimanBuzek}
\begin{eqnarray}
  \label{eq:CUnitary}
  U = \sum_{n=1}^{d^2} \ketbra{n}{n} \ten \tilde U_n. 
\end{eqnarray}

Assuming the initial state of the reservoir to be pure, that is,
$\sigma = \ketbra{\psi_0}{\psi_0}$, we obtain the phase-damping
channel
\begin{eqnarray}
  \label{eq:tqd-channel2}
  \rho'_{mn}  
  &=& \braket {\psi_n}{\psi_m} \rho_{mn}
\end{eqnarray}
in terms of the dynamical vectors $\ket{\psi_n}:= \tilde U_n
\ket{\psi_0}$, $n=1,\ldots,d^2$.

The properties of the phase-damping channel are now encoded in these
relative environment states $\ket{\psi_n}$. In particular, the
extremality of the channel is equivalent to
$\{\ketbra{\psi_n}{\psi_n}\}$ being an operator basis. A constructive
way of testing may be done using the Bloch representation. Recall that
to a given normalized complex vector $\ket{\psi_n} \in \mathds C^d$ we
can assign a corresponding generalized real Bloch vector $\vec b_n \in
\mathds R^{d^2-1}$ \cite{ZyczkowskiBook}. Let $\sigma_1, \ldots,
\sigma_{d^2-1}$ be orthogonal generators of $SU(d)$, that is, the
$\sigma_i$ are hermitian, traceless operators obeying $\tr \, \sigma_i
\sigma_j = 2 \delta_{ij}$. Together with the identity operator $\id$
these form an orthogonal basis of all linear operators in $d$
dimensions, and we arrive at the Bloch representation by defining
$\ketbra{\psi_n}{\psi_n} =: \vec B_n \cdot \vec \sigma$, where $\vec
B_n = \frac{1}{2} (\frac{2}{d},\,\vec b_n \,) \in \mathds R^{d^2}$ and
$\vec \sigma = (\id, \sigma_1, \ldots, \sigma_{r^2-1})$.  For a set of
$d^2$ projectors $\{\ketbra{\psi_n}{\psi_n} \}$ forming an operator
basis, $\{ \vec B_n \}$ is a linear independent set spanning $\mathds
R^{d^2}$.

We thus arrive at the following equivalence (see
App.~\ref{app:volume}):
\begin{eqnarray}
  \label{eq:fsov}
  &\mbox{The channel defined via the dynamical vectors }& \nonumber \\
  &\{\ket{\psi_1}, \ldots, \ket{\psi_{d^2}} \}
  \mbox{ is an extremal channel}& \nonumber \\ 
  &\Leftrightarrow&  \\
  &\mbox{Vol}(\vec b_1,\ldots,\vec b_{d^2}) \ne 0.& \nonumber
\end{eqnarray}
We are thus able to link the extremality of the phase-damping channel
to the volume $\mbox{Vol}(\vec b_1,\ldots,\vec b_{d^2}) := 1/(d^2-1)!
\; \det \left[ \begin{array}{cccc} (\vec b_2 - \vec b_1) & (\vec b_3 -
    \vec b_1) & \cdots & (\vec b_{d^2} - \vec b_1) \end{array}\right]$
spanned by the real vectors $\vec b_1,\ldots,\vec b_{d^2}$.  In this
geometric picture we can infer that the channel is extremal iff the
Bloch vectors $\vec b_n$ do not point to the same hyperplane in
$\mathds R^{d^2-1}$, or, equivalenty, iff the $d^2-1$ dimensional
volume $V$ spanned by the Bloch vectors is different from zero.

While still not a general test for the RU property, we would like to
note that, nonetheless, criterion (\ref{eq:fsov}) may be used to give
a constructive test of a channel's extremality. Given an arbitrary
phase-damping channel $D$, the Cholesky factorization gives, as
introduced in Sec.~\ref{sec:channels}, a set of dynamical vectors
$\ket{a_n} \in \mathds C^r$. Recall that $r$ denotes the rank of the
matrix $D$. Any $r^2$-dimensional subset of the corresponding Bloch
vectors $\vec b_n$ has now to be checked for linear independence. If
linear independence is found in any subset, then---following the
equivalence in (\ref{eq:fsov})---we may conclude upon extremality of
the channel. For $r \ne 1$ this immediately excludes the RU property.

\section{Symmetries and Decoherence-free Subspaces}
\label{sec:DFS}
In qubit systems it may happen that environmental influences affect
different qubits in the same way. If, for instance, the wavelength of
a fluctuating field is much larger than the separation of the qubits
certain qubit states accumulate the same random phase and coherence
among such states is preserved.  To give an example consider a
classical fluctuating magnetic field that couples identically to all
qubits via $B(t) \sum_i \sigma_z^i =: B(t) \Sigma_z$. In such a case
all superpositions of states from an eigenspace of $\Sigma_z$ will not
suffer from decoherence \cite{Lidar}. Such decoherence-free subspaces
(DFS) can be identified in experiments \cite{Schmidt-Kaler}.

In the quantum channel formalism the DFS appear naturally through
symmetry considerations. Assume, for simplicity, an N-qubit setup
where all qubits are affected by the environment in the same
way. Formally, this amounts to the invariance of the channel under
permutations of the qubits. In turn, the set of dynamical vectors
$\ket{a_n}$ has to be invariant under qubit permutations. We conclude
that $\ket{a_n} = \ket{a_m}$ whenever $\bra{n} \Sigma_z \ket{n} =
\bra{m} \Sigma_z \ket{m}$. Thus, only $N+1$ different dynamical
vectors $\ket{b_k}$ occur with a degeneracy of $\begin{pmatrix} N \\
  k \end{pmatrix}$ (the dimension of the corresponding DFS), summing
up to the total of $2^N$.

To give an example, for a two-qubit system with full qubit symmetry $1
\leftrightarrow 2$ the most general phase-damping channel is made from
only three dynamical vectors $\ket{a_1}=\ket{b_1}, \ket{a_2} = \ket{a_3}
= \ket{b_2}, \ket{a_4} = \ket{b_3}$, such that
\begin{eqnarray}
  D = \begin{pmatrix}
    1 & \braket{b_2}{b_1} & \braket{b_2}{b_1} & \braket{b_3}{b_1} \\
    \braket{b_1}{b_2} & 1 & 1 & \braket{b_3}{b_2} \\
    \braket{b_1}{b_2} & 1 & 1 & \braket{b_3}{b_2} \\
    \braket{b_1}{b_3} & \braket{b_2}{b_3} & \braket{b_2}{b_3} & 1 \\
  \end{pmatrix}
\end{eqnarray}
and the space $\{ \ket{01}, \ket{10} \}$ is a DFS. These
considerations can of course be adapted to cases of partial symmetries
of the environmental influences.

\section{Conclusions and Outlook}

We study phase damping (pure decoherence) from an environmental
perspective. Any given phase-damping channel may be understood in
terms of an overlap of dynamical vectors $\ket{a_n}$ characterizing
the channel. For a quantum environment these are relative
environmental states. We investigate how the nature of a phase-damping
process inflicts with properties of these dynamical vectors. 

For a single qubit, we infer that any possible phase-damping channel
is indeed Markovian, that is, a snapshot of some time-continuous
Markovian evolution. For a single qutrit, we find examples of channels
that are not Markovian: we give a class of channels we show is not
infinitely divisible. Remarkably, it turns out that in case of
Markovian phase damping in arbitrary dimension the dynamical vectors
may be chosen to be multi-mode coherent states.

For a single qubit a phase-damping channel is of RU type. For Hilbert
space dimension $N \geq 4$ we discuss a physical model of
phase-damping dynamics that has no RU representation. We find that for
a phase-damping channel acting on a $d^2$ dimensional quantum system,
the RU property may be linked to a $(d^2-1)$-dimensional volume. In a
previous article, a link between this (absolute) volume and the norm
distance between the channel and the convex hull of unitary
transformations was found \cite{HelmStrunz}.

Our considerations are of relevance for process tomography
\cite{NielsenChuang, JuanPabloPaz} where it is a great challenge to
reduce the dimension of the parameter space of the process.  It is
clear that any additional assumption about the nature of the process
(phase damping, RU, Markovian) leads to further constraints.  Our
results allow for a characterization of the channel with a minimal
number of parameters and should help to speed-up the optimization
procedures involved \cite{James}.

\section*{Acknowledgments}
W.T.S. is grateful to Thomas Seligman and the Centro Internacional de
Ciencias in Cuernavaca, Mexico, where part of this work took shape.
J.H. acknowledges support from the International Max Planck Research
School (IMPRS) Dresden.

\appendix
\section{Diagonal Lindblad Form}
\label{app:diagonality}
In this appendix we show that the diagonal form of the generator
$\mathcal L$ is enough to guarantee the Hamiltonian and the Lindblad
operators to be diagonal as well. In a given basis $\{\ket{n}\}$, let
the generator of the semigroup $\Lambda_t = e^{\mathcal L t}$ be
diagonal, that is, $\mathcal L(\ketbra{m}{n}) = z_{mn}
\ketbra{m}{n}$. With $H = \sum_{mn} h_{mn} \ketbra{m}{n}$ and $L_i =
\sum_{mn} l_{mn}^{(i)} \ketbra{m}{n}$ this implies
\begin{widetext}
  \begin{eqnarray}
    \label{appeq:lindblad}
    z_{mn} \delta_{rm}\delta_{ns} &=& 
    -i \left(h_{rm} \delta_{ns}-\delta_{rm} h_{ns} \right) 
    + \frac{1}{2} \sum_{i} 
    \left\{ 
      \Big( l^{(i)}_{rm} l^{(i)}_{ns}
      - \sum_k l^{(i)}_{rk} l^{(i)}_{km} \delta_{ns} \Big) \right. 
    \left. + \Big( l^{(i)}_{rm} l^{(i)}_{ns} 
      - \sum_k l^{(i)}_{nk} l^{(i)}_{ks} \delta_{rm} \Big)
    \right\} \nonumber \\
    &=& -i \left( h_{rm} \delta_{ns} - \delta_{rm} h_{ns} \right) 
    + \braket{l_{rm}}{l_{ns}} 
    - \frac{1}{2} \sum_k \Big( \braket{l_{rk}}{l_{km}} \delta_{ns} + 
    \braket{l_{nk}}{l_{ks}} \delta_{rm} \Big). 
  \end{eqnarray}
\end{widetext}
Letting $m=n$, $r=s$, and $n \ne s$ we see that $ \| l_{mr} \|^2 = 0
\quad \mbox{for} \quad m \ne r$, so that $l_{mr} = \delta_{mr}
l_r$. Insertion into (\ref{appeq:lindblad}) then implies $h_{mr} =
\delta_{mr} \omega_r$, so that Hamiltonian and Lindblad operators are
found to be diagonal. In matrix representation, the generator may thus
be written as
\begin{eqnarray*}
  z_{mn} = -i(\omega_m - \omega_n) + \braket{l_n}{l_m} 
  - \frac{1}{2} (\|l_m\|^2 + \| l_n\|^2). 
\end{eqnarray*}

\section{Extremality Criterion}
\label{app:volume}
In order to see the equivalence in Eq.~(\ref{eq:fsov}) we have to
perform some matrix algebra. The vectors
$\ket{\psi_1},\ldots,\ket{\psi_{d^2}}$ define projectors giving an
operator basis iff the real vectors $\vec B_1,\ldots,\vec B_{d^2}$ are
linearly independent, which is the case for \cite{HornJohnson}
\begin{eqnarray*}
    && 
    \det
    \left(
      \begin{array}{cccc} 
        {} \\
        \vec B_1 & \cdots & \vec B_{d^2} \\
        {}
      \end{array}
    \right) =
    \det
    \left(
      \begin{array}{cccc} 
        \frac{2}{d}&\cdots&\frac{2}{d}\\{}\\
        \vec b_1 & \cdots & \vec b_{d^2} 
      \end{array}
    \right) \\
    &=& 
    \det
    \left(
      \left(
        \begin{array}{cccc} 
          \frac{2}{d}&\cdots&\frac{2}{d}\\{}\\
          \vec b_1 & \cdots & \vec b_{d^2}  \\ {} 
        \end{array}
      \right)
      \left(
        \begin{array}{ccccc} 
          1 & -1 & \cdots & -1\\
          & 1  & &   \\
          &    &  \ddots & \\
          & & & 1
        \end{array}
      \right)
    \right) \nonumber \\
    &=& \frac{2}{d} \det
    \left(
      \begin{array}{cccc} 
        (\vec b_2 - \vec b_1) & (\vec b_3 - \vec b_1) & \cdots & (\vec b_{d^2} - \vec b_1)
      \end{array}
    \right) \nonumber \\
    &=& \frac{2 (d^2-1)!}{d} \; \; \mbox{Vol}(\vec b_1,\ldots,\vec b_{d^2}) \\
    &\ne & 0, 
  \end{eqnarray*}
  where $\mbox{Vol}(\vec b_1,\ldots,\vec b_{d^2})$ denotes the volume
  of the parallelogram spanned by the real, $d^2-1$ dimensional
  vectors $\{ \vec b_1,\ldots,\vec b_{d^2} \}$.

\bibliography{mybib}{}

\begin{thebibliography}{99}
\bibitem{Giulini}
  E. Joos {\it et al.}, {\it Decoherence and the Appearance of a
    Classical World in Quantum Theory}, 2nd ed. (Springer, New York,
  2003).

\bibitem{ZurekReview} 
  W. H. Zurek, Rev. Mod. Phys. {\bf 75}, 715 (2003).

\bibitem{StrunzDecoherence} 
  W.T. Strunz, {\it Decoherence in Quantum Physics} in {\it Coherent
    Evolution in Noisy Environments}, A. Buchleitner and
  K. Hornberger, Eds. (Springer Lecture Notes in Physics vol. 611,
  Berlin, 2002).

\bibitem{Yu2003} 
  T. Yu and J.H. Eberly, Phys. Rev. B {\bf 68}, 165322 (2003).

\bibitem{BreuerBook} 
  H.-P. Breuer and F. Petruccione, {\it The Theory of Open Quantum
    Systems} (Oxford University Press, Oxford, 2002).

\bibitem{Breuer2009}
  H.P. Breuer, Phys. Rev. Lett. {\bf 103}, 210401 (2009).

\bibitem{Wolf2008a} 
  M.M. Wolf and J.I. Cirac, Commun. Math. Phys. {\bf 279},
  147--168 (2008).

\bibitem{Wolf2008b} 
  M.M. Wolf, J. Eisert, T.S. Cubitt and J.I. Cirac, 
    Phys. Rev. Lett. {\bf 101}, 150402 (2008).

\bibitem{AlickiLendi} 
  R. Alicki and K. Lendi, {\it Quantum Dynamical Semigroups and
    Applications} (Springer, New York, 1987).

\bibitem{NielsenChuang}
  M.A. Nielsen and I.L. Chuang, {\it Quantum Computation and Quantum
    Information} (Cambridge University Press, Cambridge, UK, 2007).

\bibitem{LandauStreater} 
  L.J. Landau and R.F. Streater,  Linear Algebr. Appl. {\bf 193},
  107 (1993).

\bibitem{DAriano}
  F. Buscemi, G. Chiribella, and G. M. D'Ariano, 
    Phys. Rev. Lett. {\bf 95}, 090501 (2005).

\bibitem{HelmStrunz}
  J. Helm and W.T. Strunz, Phys. Rev. A {\bf 80}, 042108 (2009).

\bibitem{ZyczkowskiBook}
  I. Bengtsson and K. \.{Z}yczkowski, {\it Geometry of Quantum States}
  (Cambridge University Press, Cambridge, UK, 2006).

\bibitem{Denisov}
  L.V. Denisov, Theor. Prob. Appl. {\bf 33}, 392 (1988).

\bibitem{GregorattiWerner}
  M. Gregoratti and R.F. Werner, J. Mod. Opt. {\bf 50}, 915 (2003).

\bibitem{Alber}
  J. Novotny, G. Alber, and I. Jex, J. Phys. A. {\bf 42}, 282003 (2009).

\bibitem{Havel}
  T.F. Havel {\it et al.}, Phys. Lett. A {\bf 280}, 282 (2001).

\bibitem{2004GorinStrunz}
T. Gorin, T. Prosen, T. H. Seligman, and W. T. Strunz,
Phys. Rev. A {\bf 70}, 042105 (2004).

\bibitem{Gentle} 
  Gentle, J.E., {\it Numerical Linear Algebra for Applications in
    Statistics} (Springer, Berlin 1998).

\bibitem{ScullyBook}
  M.O. Scully and M.S. Zubairy, {\it Quantum Optics} (Cambridge
  University Press, Cambridge, UK, 1997).

\bibitem{ZimanBuzek}
  M. Ziman and V. Bu\v{z}ek, Phys. Rev. A {\bf 72}, 022110 (2005).

\bibitem{Lidar} 
  D.A. Lidar and K.B. Whaley, {\it Decoherence-Free Subspaces and
    Subsystems} in {\it Irreversible Quantum Dynamics}, F. Benatti and
  R. Floreanini, Eds. (Springer Lecture Notes in Physics vol. 622,
  Berlin, 2003).

\bibitem{Schmidt-Kaler} 
  H. H\"{a}ffner, F. Schmidt-Kaler, C.F. Roos, T., K\"{o}rber,
  M. Chwalla, M. Riebe, J. Benhelm, U.D. Rapol, C. Becher, and
  R. Blatt, Appl. Phys. B {\bf 81} 151 (2005).

\bibitem{JuanPabloPaz} 
  A. Bendersky, F. Pastawski, and J.P. Paz, Phys. Rev. Lett. {\bf
    100}, 190403 (2008).

\bibitem{James} 
  D.F.V. James, P.G. Kwiat, W.J. Munro, and A.G. White, Phys. Rev. A
  {\bf 64}, 052312 (2001).

\bibitem{HornJohnson}
  R.A. Horn and C.R. Johnson, {\it Matrix Analysis} (Cambridge
  University Press, Cambridge, New York, 2007).

\end{thebibliography}

\end{document}